\documentclass[journal ]{IEEEtran}
\usepackage[utf8]{inputenc}
\usepackage{textcomp}

\usepackage{graphicx}
\usepackage{comment}
\usepackage{amsmath}
\usepackage{amssymb}
\usepackage[version=4]{mhchem}
\usepackage{siunitx}
\usepackage{longtable,tabularx}
\usepackage{xcolor}
\newcommand{\change}[1]{{\color{purple} \textbf{#1}}}
\renewcommand{\change}[1]{#1}

\title{Graph Percolation as Decision Threshold for Risk Management in Cross-Country Thermal Soaring}

\author{John J. Bird,~\IEEEmembership{Member,~IEEE}
\thanks{J. Bird was with the Department of Aerospace and Mechanical Engineering, The University of Texas at El Paso,
El Paso, TX, 79902 USA e-mail: (see https://hb2504.utep.edu/Home/Profile?username=jjbird).}}

\begin{document}

\section*{IEEE Copyright Notice}
\noindent©2026 IEEE. Personal use of this material is permitted. Permission from IEEE must be obtained for all other uses, in any current or future media, including reprinting/republishing this material for advertising or promotional purposes, creating new collective works, for resale or redistribution to servers or lists, or reuse of any copyrighted component of this work in other works.

\vspace{1em}
\noindent Accepted for publication in IEEE Transactions on Aerospace and Electronic Systems

\newpage

\maketitle

\begin{abstract}
Long range flight by fixed-wing aircraft without propulsion systems can be accomplished by ``soaring'' -- exploiting randomly located updrafts to gain altitude which is expended in gliding flight. As the location of updrafts is uncertain and cannot be determined except through in situ observation, aircraft exploiting this energy source are at risk of failing to find a subsequent updraft. Determining when an updraft must be exploited to continue flight is essential to managing risk and optimizing speed. Graph percolation offers a theoretical explanation for this risk, and a framework for evaluating it using information available to the operator of a soaring aircraft in flight. The utility of graph percolation as a risk measure is examined by analyzing flight logs from human soaring pilots. This analysis indicates that in sport soaring pilots rarely operate in a condition which does not satisfy graph percolation, identifies an apparent desired minimum node degree, and shows that pilots accept reduced climb rates in order to maintain percolation.
\end{abstract}

\section{Introduction}
\begin{figure}
    \centering
    \includegraphics[width=\linewidth]{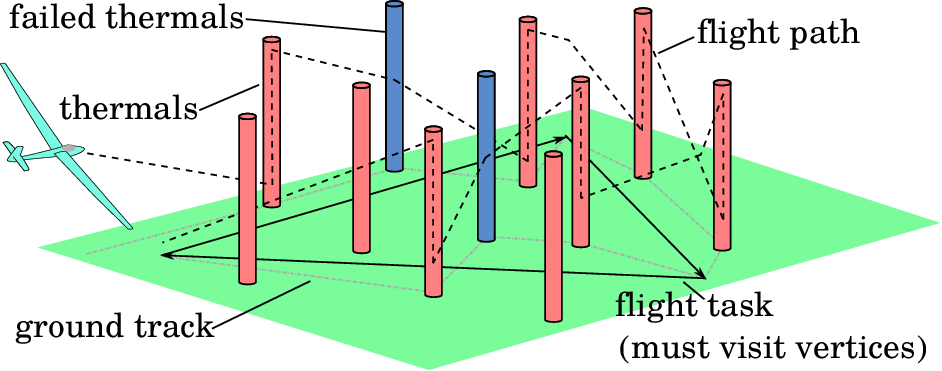}
    \caption{Cross-country soaring requires exploiting a sequence of updrafts to gain energy which is expended in gliding flight. Updrafts are stochastically distributed and cannot be reliably detected except in situ, putting an aircraft at risk of failing to find additional energy sources before reaching the ground.}
    \label{fig:soaring}
\end{figure}

In thermal soaring an unpowered aircraft sustains flight by circling within updrafts called ``thermals'' where the vertical air motion, or ``lift,'' exceeds the aircraft's unpowered descent rate.
The aircraft thus climbs, gaining energy which is stored as altitude then expended in a glide to reach the next thermal or an objective point\cite{Rei81} as seen in figure \ref{fig:soaring}.
Soaring is practiced by birds to provide economical locomotion, by humans seeking to maximize speed as a competitive sport\cite{Rei81}\cite{Brigliadori2006}, and by Uncrewed Aircraft Systems (UAS) to extend range and endurance\cite{Walton2017}\cite{Edwards2010}\cite{Allen2007}\cite{Depenbusch2017b}.

Classic thermal cross-country soaring theory focuses on optimizing the speed between lift sources of known strength, under the assumption that there is adequate altitude to reach the next thermal\cite{Rei81}.
Under these assumptions, ``speed-to-fly theory'' can be derived which clearly defines the optimal speed in response to random air motion and expected thermal strength for a given aircraft.
As the strength of expected lift increases a faster speed is optimal -- the greater descent rate which occurs at higher speeds is compensated by more rapid climbs to regain energy.
The expected next thermal strength is often referred to at the ``MacCready'' number after an instrument developed by Paul MacCready which provided an instantaneous indication of the optimal speed to fly\cite{MacCready1958}.
The stochastic nature of thermal formation poses a challenge to this model however. 
Thermals are not uniformly spaced or of uniform strength, so a pilot may occasionally have to exploit weaker thermals in order to continue flight as failure to find a thermal results in an ``outlanding'' in which the aircraft must land in an improvised landing area (e.g. an agricultural field).
In a competition an outlanding results in a poor score and can effectively place a pilot out of contention for top rankings.

In the soaring community the interaction of sporting risk and speed is acknowledged in a few ways -- pilots often speak of a ``working band'' of altitude.
When the altitude is maintained within this band the mean speed is relatively fast\cite{Brigliadori2006}. 
Pilots may also speak of ``shifting gears'' when encountering difficult conditions or missing a climb, into a mindset where they will accept weaker climbs to ``reconnect'' with good conditions\cite{Bird2019b}. 
Efforts to formalize risk management have been made, most notably by Fukada\cite{Fukada2000} and Cochrane\cite{Cochrane1999} who studied the impact that speed has on expected competition score when lift sources are not assured. 
These analyses show that as a pilot's altitude decreases they should slow down and accept lower climb rates to hedge against not having enough energy to find the next thermal. 
This can be expressed in the MacCready setting that the pilot chooses to guide their speed to fly.

This use of the MacCready setting is intuitive to soaring pilots, but does not directly address risk in a theoretically rigorous way.
An alternative is to model cross country soaring as a random graph traversal. 
A flight takes place between nodes (thermals) which are connected by edges if one node can be reached from another (the thermals are within gliding range). 
Because of the stochastic nature of thermal formation the locations of the nodes (and thus their interconnections) are random. 
The possibility of completing a sequence of waypoints (a ``cross-country task'') can be stated in terms of the existence of a path which connects a sequence of nodes that traverse the task.
While not directly answering this question, continuum percolation allows properties of paths on random graphs to be determined\cite{Gilbert1961}, and indeed this has been applied to thermal soaring in the past\cite{Depenbusch2014}. 
Percolation suggests that once enough nodes are reachable from a given node that the probability of continued cross-country flight suddenly approaches 1 (though not necessarily along the desired task). 
Indeed, Depenbusch verified this behavior in Monte Carlo simulation and used percolation analysis to derive constraints on aircraft performance as a function of thermal spacing\cite{Depenbusch2014}. 

The present work takes percolation analysis a step further. By considering a slightly different percolation problem, operational (rather than design) bounds on soaring performance can be obtained. 
This manuscript briefly reviews continuum percolation as applied to cross-country soaring, suggests a new percolation condition, and applies a post hoc percolation analysis to flights completed by human pilots to assess the utility of percolation as a risk measure.
The results suggest that percolation offers an effective means to assess soaring risk and a path to unify risk management and speed to fly.
This has the potential to improve cross-country soaring practiced as a sport by human pilots, and also to provide a principled basis for autonomous decision making when soaring is used to defray energetic costs for UAS.

\section{Graph Percolation and Soaring}
\begin{figure}
    \centering
    \includegraphics[width=\linewidth]{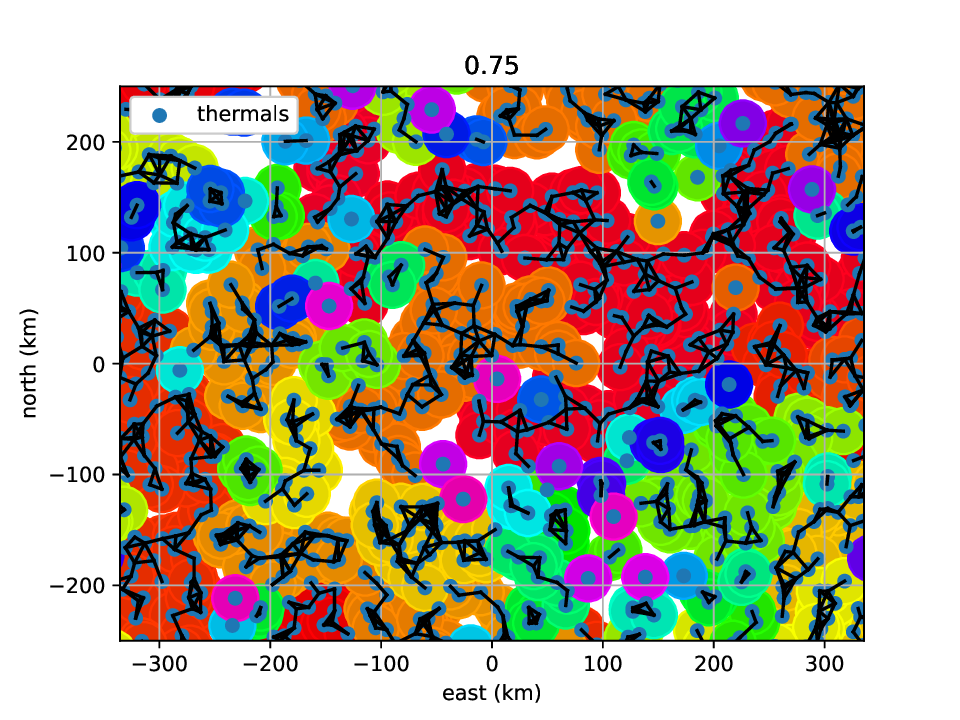}
    \caption{A collection of thermals which forms a random geometric graph whose connection degree is three-quarters of the critical threshold. The colored area surrounding each thermal indicates the glide range possible from the top of the thermal. Colors indicate clusters of thermals which can be traversed.}
    \label{fig:random_graph}
\end{figure}

Continuum percolation is a property of random geometric graphs, which are composed of nodes or vertices which are randomly located and connected to each other if some condition is satisfied. 
Figure \ref{fig:random_graph} illustrates a thermal graph $\mathcal{G}$: vertices $\mathcal{V}$ are located at thermals whose locations are generated by a Poisson process, and connected with an edge $\mathcal{E}$ if a thermal is within gliding range of the previous thermal.
The graph is colored by traversable clusters (every thermal in a traversable cluster can be reached from at least one other thermal in the cluster). 
Note that in general edges do not need to be two-way connected, but for simplicity we will consider symmetric edges (equivalent to constant thermal height and no wind).

If the graph is fully known then cross-country soaring can be solved as a graph planning problem. 
However in soaring, the graph is not only random, but unknown.
Thermals form and dissipate with a characteristic time of approximately 15 minutes\cite{Stull1988} and cannot in general be reliably detected except by in situ observation.
The gross statistics of a thermal graph is thus more suitable for analyzing cross-country soaring than is direct graph traversal. 
Continuum percolation considers the structure and long-range connectedness of random graphs which exist in an infinite, homogeneous space. 
When considering the number of thermals in a traversable cluster, percolation theory indicates that as the average number of thermals reachable from any given thermal increases above a critical threshold that a phase change occurs in the graph. 
Below this threshold, the size of (number of thermals in) the largest cluster is finite. 
Upon crossing this threshold, with probability approaching 1 a cluster exists in the graph which contains an infinite number of thermals.

\begin{figure}
    \centering
    \includegraphics[width=\linewidth]{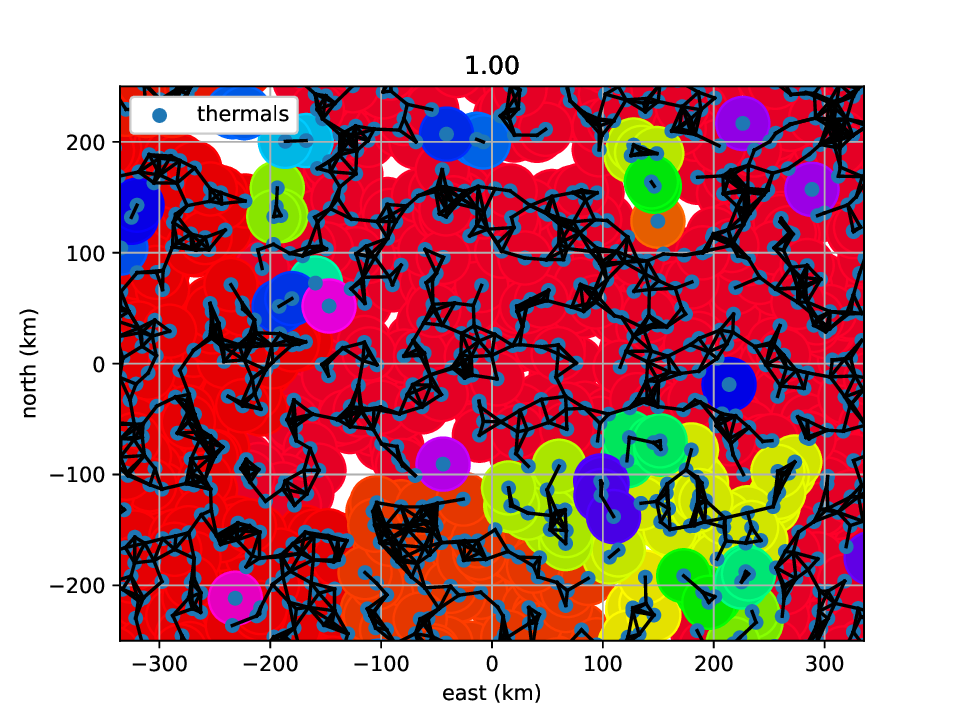}
    \caption{A collection of thermal which forms a random geometric graph whose connection degree is at the critical threshold. The colored area surrounding each thermal indicates the glide range possible from the top of the thermal. Colors indicate clusters of thermals which can be traversed.}
    \label{fig:random_graph_percolating}
\end{figure}

Figure \ref{fig:random_graph_percolating} shows a graph formed by the same thermals, but with a longer gliding range (e.g. higher altitude) so that the graph is exactly at the percolation threshold.
Most of the graph is now covered by a single large cluster. 
\change{Percolation as an enabler of cross-country soaring rests on a probabilistic argument -- that if there are clusters with infinite numbers of thermals in them, then if a thermal is selected at random it probably lies in one of these clusters. 
This sudden phase change in the graph appears consistent with soaring experience -- as the aircraft performance or environmental conditions cross a threshold suddenly cross-country soaring becomes relatively ``easy.''
Note however that this does not guarantee that any given location can be reached from a cluster, that a flight objective lies in a cluster, or that the path between two points in a cluster is particularly short or efficient.}

\change{A key question to answer is: what is the critical degree of each node, that is, on average how many thermals must be reachable to ensure the graph percolates? While theoretical bounds on this threshold exist, they are too loose to be of practical use, giving the critical degree as between 1.75 and 10.9\cite{Gilbert1961}. Monte Carlo simulations conducted by constructing large random graphs and analyzing their statistics give the critical degree as lying between 4.508 and 4.515\cite{Balister2005}. Here we adopt 4.51 as the threshold; when the expected number of thermals reachable from a given point is 4.51 then the point is said to ``percolate'' and it is anticipated the cross country soaring is possible from this point.}

Depenbusch applied percolation analysis to soaring flight in an attempt to derive a bound on the aircraft performance required for cross-country flight\cite{Depenbusch2014}. 
This analysis indicated that for realistic thermal densities that the required glide ratio (range achievable per unit of altitude) to achieve percolation was only 5-15. 
Initially this seems like a very loose bound (modern sailplanes have glide ratios often exceeding 50), but birds and paragliders with performance in this range regularly practice cross-country soaring\cite{Pennycuick1972}\cite{Becker2017}.


If the locations of thermals are modeled as generated by a spatial Poisson point process with intensity $\lambda_2$ representing the expected number of thermals per unit area, then the expected number of thermals reachable from any point was derived by Depenbusch\cite{Depenbusch2014}:
\begin{equation}
    \mathbb{E}(n) = \lambda_2 \pi r^2 = \lambda_2 \pi \left(h\frac{L}{D}\right)^2
    \label{eq:depenbusch_percolation}
\end{equation}
where $r$ is the gliding range of the aircraft, $h$ is its remaining usable altitude, and $\frac{L}{D}$ is the aircraft's glide ratio. 
The distinction between this analysis and that of Depenbusch lies in the choice of $h$ -- Depenbusch chooses $h$ to be the altitude usable from the top of one thermal before reaching the ground. 
Here we choose to evaluate percolation at every instant in flight according to the aircraft's current usable altitude. 
While in principle it would be acceptable to have fewer than the threshold number of thermals in range at some points in flight (e.g. at the end of a glide as the aircraft begins to climb in the next thermal), we observe that because it is not known when the next thermal will be available (until the next climb actually begins) that the threshold connection should be maintained at all times.

The ``percolation value'' can then be defined as the ratio of the expected number of thermals reachable (the degree of a node representing the aircraft's location) to the critical node degree:
\begin{equation}
\begin{aligned}
    p =&\\ 
    =&\frac{\mathbb{E}(n)}{n_{critical}} \\
    =& \lambda_2 \pi \left(h\frac{L}{D}\right)^2\frac{1}{n_{critical}} \\
    \approx& \lambda_2 \pi \left(h\frac{L}{D}\right)^2\frac{1}{4.51}
\end{aligned}
\end{equation}
Most of this expression is relatively straightforward to define, except for the intensity of the thermal location process. 
If the thermals are generated by a Poisson process, then it represents the expected number of thermals \textit{per unit area} which is difficult to determine. 
The intensity of a one-dimensional process describing the thermals encountered by the aircraft along its route of flight can be determined however, simply by counting the number of thermals and integrating the flight path length.
It can be shown that the expected distance from any point in the plane to a point generated by a spatial Poisson process with intensity $\lambda_2$ (i.e. a thermal) can be described (appendix \ref{sec:intensity}):
\begin{equation}
    \mathbb{E}(r) = \frac{1}{2\sqrt{\lambda_2}}
    \label{eqn:expected_spacing}
\end{equation}
because the spatial Poisson process has the same statistics for any point in the plane, the expected distance to the nearest point holds for any point selected. 
Considering the location of the aircraft at any time, the expected distance to the next thermal is given by Equation \ref{eqn:expected_spacing} which is inverse of the intensity parameter, $\lambda_1$ in a point process for the distance between nearby thermals. 
Thus, the intensity of the process describing distance between thermals and that describing the spatial intensity is given:
\begin{equation}
    \frac{1}{\lambda_1} = \frac{1}{2\sqrt{\lambda_2}}
\end{equation}
which can be rearranged
\begin{equation}
    \lambda_2 = \left(\frac{\lambda_1}{2}\right)^2
    \label{eq:lambda_1_to_2}
\end{equation}
to determine the spatial intensity of thermals.

Combining equations \ref{eq:lambda_1_to_2} and \ref{eq:depenbusch_percolation}, the ``percolation'' value can be defined:
\begin{equation}
    p \approx \left(\frac{\lambda_1}{2}\right)^2 \pi \left(h\frac{L}{D}\right)^2\frac{1}{4.51}
\end{equation}
where $\lambda_1$ can be approximated from the observed distance between thermals. 
When $p$ is greater than 1 the present altitude, thermal intensity, and lift-to-drag ratio is such that sufficient additional thermals are reachable so that the graph of thermal connections percolates and indefinite cross-country flight is expected to be possible (though again, it is not guaranteed that the aircraft position or objective are actually in a large cluster). 
Thus, the percolation value is interesting both as a diagnosis of risk taken in previous flights and with an eye to the development of future soaring instrumentation and strategies.

\section{Flight Data Analysis}
Soaring pilots regularly log their flights to enable competition scoring and to share as a social experience\cite{WeGlideUG2025}.
These logs provide a rich source of data to test the utility of percolation.
Unfortunately, the data in flight logs is fairly limited, typically containing only position and time information.
Even given the relationship in equation \ref{eq:lambda_1_to_2}, defining the thermal intensity is a challenge.
Thermals are only readily identified in flight logs from the circling maneuvers used to exploit them, so any thermals which a pilot does not exploit will not be identified, reducing the apparent thermal density.
Further, comparing flights across days requires defining thermal intensity for each day.
Some of these challenges can be resolved by examining logs from contest flights where many pilots operate in a similar area at the same time.

\subsection{Contest Flight Analysis}
Flight logs were obtained for the 2024 United States Club Class National Championship which took place in Hobbs, New Mexico in June of 2024.
The topography surrounding the contest site is flat and conditions tend be be relatively uniform with strong thermal updrafts.
Contest tasks each day typically lasted three hours and covered roughly 300 km distance. 
As this is a national championship competition, it is assumed that pilots are generally skilled and will exploit the fewest number of the strongest thermals possible in order to achieve high speeds.
We hypothesize that the pilots will attempt to stay above the percolation threshold at all points in flight.
We also assume that at least one pilot will exploit every ``good'' thermal (thermals with above-average but not exceptional updraft velocity) they encounter, allowing the intensity of the process generating ``good'' thermals to be defined by the closest observed thermal spacing for a given day.
In principle this is not a competition optimal strategy, but in the author's experience flying sailplane competitions, the range of pilot expertise is such that this is approximately satisfied.

While thermal strength and distribution varies through the day, contest flying generally occurs during the afternoon with relatively constant conditions, so a single thermal intensity is applied to all flights on a given day. 
Assuming that pilots stop attempting to exploit thermals in order to focus on a successful outlanding at a height of 200 m above ground, and that club class gliders achieve a glide ratio in cruise of $\frac{L}{D}=30$\cite{Rohde-Brandenburger2017}, then percolation can be computed for every point in the flight, with special emphasis given to the percolation value at the moment where thermalling begins.
Pilots adopt different strategies prior to starting a task and when finishing it, so every point within 10 km of takeoff is removed, as is the flight segment following the last thermal\cite{Rei81}. 
One of the contest days is discarded due to thunderstorm activity which resulted in a sudden change in the soaring conditions during the task.


\begin{figure}
    \centering
    \includegraphics[width=\linewidth]{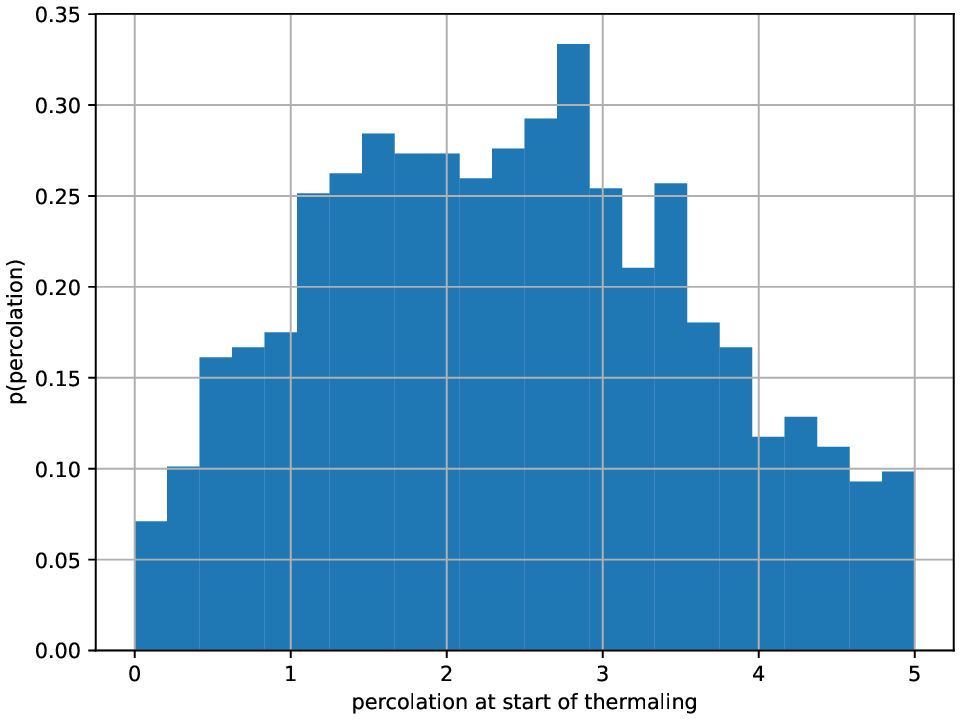}
    \caption{Histogram of the percolation value at which pilots in the Hobbs contest begin thermalling.}
    \label{fig:hobbs_histogram}
\end{figure}

Figure \ref{fig:hobbs_histogram} shows the distribution of the percolation at the start of thermalling for every thermal used by competitors during the Hobbs club class contest.
The mean of this distribution is 2.5, fewer than 15\% of all thermalling events begin at a percolation value lower than 1.0.

\begin{figure}
    \centering
    \includegraphics[width=\linewidth]{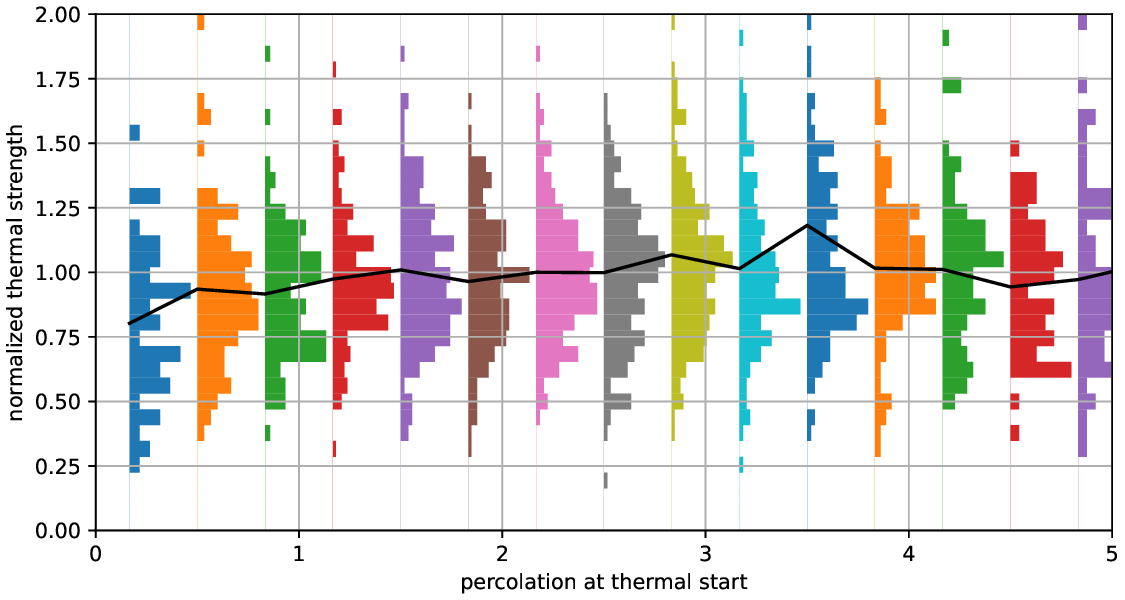}
    \caption{Normalized thermal strength distribution vs percolation at start of thermalling in Hobbs data.}
    \label{fig:hobbs_strength}
\end{figure}

While the number of outlandings is too small to evaluate the effect that percolation has on controlling this risk, the effect on speed can be approximated by examining thermal strength (stronger thermals enable faster speeds as less time is spent circling). 
Figure \ref{fig:hobbs_strength} illustrates the distribution of thermal strength (normalized by a pilot's mean thermal strength on a given day) associated with the percolation value at which thermal exploitation begins.
While the data is noisy it appears that pilots accept weaker climbs when thermalling at low percolation values, indicating that they are responding to risk of landing out.

\begin{figure}
    \centering
    \includegraphics[width=\linewidth]{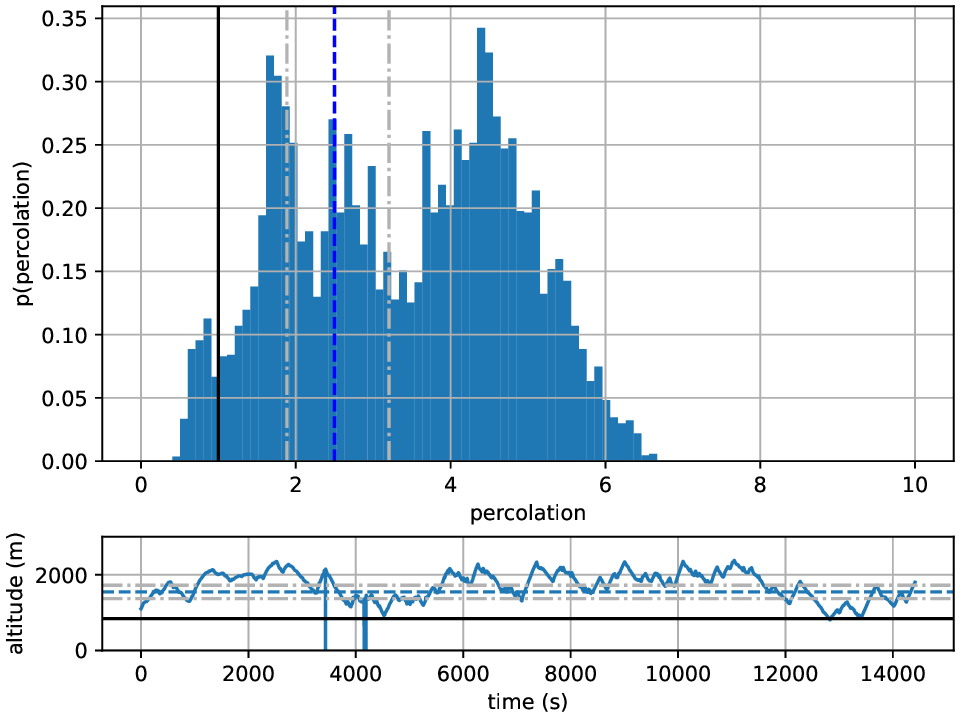}
    \caption{Percolation histogram and time history of altitude for the winner of one contest day at the Hobbs contest.}
    \label{fig:hobbs_percolation}
\end{figure}

Examining a single flight, figure \ref{fig:hobbs_percolation} shows the distribution of the percolation value for the winner on one of the competition days as well as the time history of altitude.
Note that this shows percolation at all points in the flight, distinct from the distribution illustrated in figure \ref{fig:hobbs_histogram} which shows percolation only at the start of each thermalling segment.
A percolation value of 1 is indicated by the black line and 2.5 is indicated by the blue dashed line.
Grey dash-dot lines indicate the altitude (and percolation value) at which the pilot can glide a horizontal distance of one-half of the thermal spacing before reaching $p=2.5$, and the altitude (and percolation value) the pilot would be at if they glided a horizontal distance of one-half of the thermal spacing after reaching $p=2.5$. 
This can be thought of as the altitude band we would expect the pilot to cover if they intend on average to find the next thermal at the altitude which achieves $p=2.5$.
The figure indicates that almost all thermals are encountered before falling below $p=1$, and that many thermals are encountered in the band around $p=2.5$.

\subsection{Large Database Analysis}
The Hobbs contest data indicates that pilots typically start thermalling before reaching the percolation threshold, and identifies a decision threshold that pilots appear to be using, lending credence to percolation as a useful metric for assessing risk and the necessity of exploiting a thermal.
Evaluating the effect of percolation on speed is more challenging however.
When thermals are binned according to starting percolation value, the number of thermals in each bin is small and resulting distributions are noisy.
To evaluate this effect more clearly, a large dataset of cross-country soaring flights was obtained from the flight log sharing service WeGlide, covering a flat region in central Europe.
Each flight in the set consists of a sequence of thermals visited during the flight which are defined by mean strength as well as initial and final position, altitude, and time.
In principle the intensity of thermals along the flight path for each flight can thus be approximated by the sum of the distance between thermals and the number of thermals encountered:
\begin{equation}
    \lambda_1 \approx f \frac{n_{thermals} - 1}{\sum_{i=1}^{n_{thermals}-1}D(i+1, i)}
    \label{eqn:lambda_est}
\end{equation}
where $f$ is the fraction of thermals encountered which are used and $D(i,j)$ represents the distance between thermal $i$ and thermal $j$. The fraction $f$ is needed because the dataset only includes thermals which the pilot actually exploited, so any thermals which were encountered and were not needed at the time will not appear in the dataset. 
Unfortunately the fraction $f$ cannot be determined from the available data. 
In contrast to the contest data, in general there are not many gliders flying the same route at approximately the same time. 
To resolve this, the thermal intensity is instead set so that the mean percolation value at the start of thermalling for each flight is equal to that observed in the Hobbs data. 

\begin{figure}
    \centering
    \includegraphics[width=\linewidth]{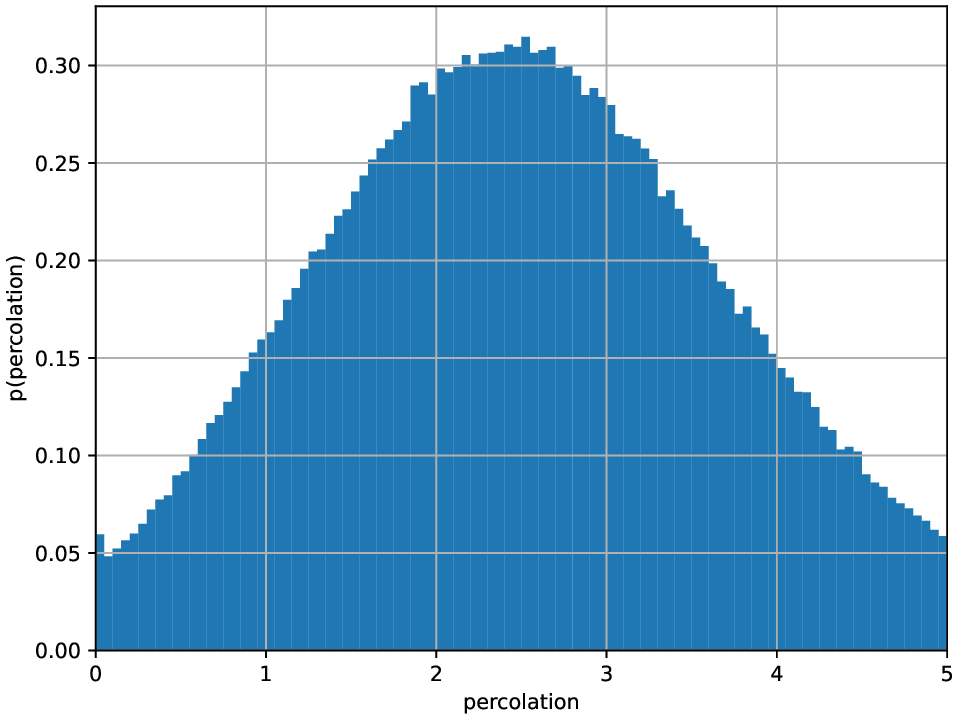}
    \caption{Normalized histogram of percolation at start of thermalling from a set of more than 350 000 thermals.}
    \label{fig:weglide_percolation}
\end{figure}

The percolation parameter can then be computed at the point where each thermalling segment begins and used to assess whether it was critical from an outlanding risk management perspective to exploit that thermal. 
We hypothesize that as percolation decreases that pilots will accept weaker thermals in order to avoid falling below the threshold and becoming ``disconnected'' from clusters of thermals. 
Assuming again that pilots reserve 200 m of altitude to accomplish an outlanding and that pilots expect to maintain a lift to drag ratio of 30, percolation can be computed at the start of each thermalling event:
\begin{equation}
    p = \left(\frac{\lambda_1}{2}\right)^2 \pi \left((h_{AGL} - \mathrm{200\ m})30\right)^2\frac{1}{4.51}
\end{equation}

This analysis is conducted for 350 218 thermals drawn from records of 14 584 flights. Figure \ref{fig:weglide_percolation} shows a histogram of the percolation value at the beginning of thermalling for each thermal, analogous to figure \ref{fig:hobbs_histogram}.
The shape overall is similar to that observed in the Hobbs dataset. 
Since the thermal spacing was derived assuming that thermalling starts on average at a percolation value of 2.5, the center of this distribution is near 2.5 as expected.
For the WeGlide dataset fewer than 10\% of all thermalling events begin at a percolation less than 1.0.

Directly assessing the effect of percolation on speed is difficult as the dataset covers many different gliders of varying performance.
The relationship between percolation and thermal strength can be examined as a proxy (stronger thermals should be associated with faster achieved cross-country speeds as pilots gain energy more quickly).
The strength of each thermal is extracted and normalized by the mean thermal strength for each flight.
Thermals are binned by the percolation value at the start of thermalling, with bin sizes of 0.1. 
Statistics of each bin are computed and used to establish the relationship between percolation at start of thermalling and thermal strength. 

\begin{figure}
    \centering
    \includegraphics[width=\linewidth]{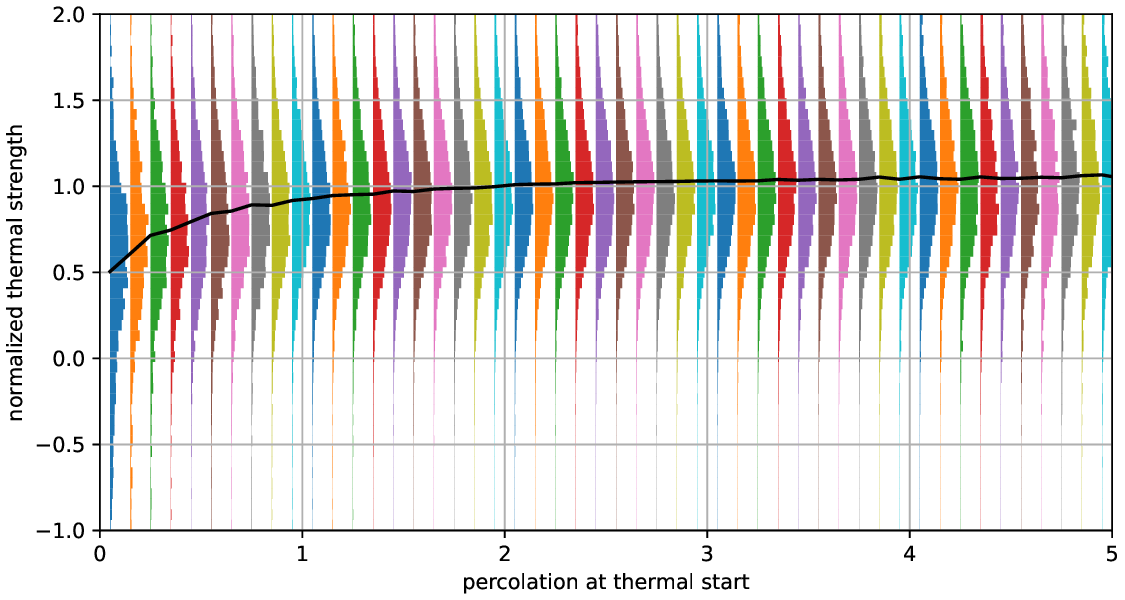}
    \caption{Normalized thermal strength distribution vs percolation at start of thermalling in WeGlide data.}
    \label{fig:weglide_thermal_strength}
\end{figure}

Figure \ref{fig:weglide_thermal_strength} shows this relationship, with histograms of thermal strength for each bin and a line indicating the mean value for each bin. 
When pilots begin thermalling at a percolation value of 2.0 or greater they use thermals at or above their flight average. As the percolation value decreases, and especially as it falls below 1, pilots use weaker and weaker thermals. 
It is also evident that as percolation value falls pilots are more likely to attempt to exploit thermals which do not result in a climb at all (negative strength thermals), attempting to exploit very weak thermals which may not result in an altitude gain. 

\section{Discussion}

Random graphs and their properties appear to be a useful tool in modeling and analyzing cross-country soaring flight.
Theoretical properties of percolation on random 2D graphs explains the qualitative experience in cross-country soaring that a pilot can suddenly find it much more challenging to stay aloft.
By considering the information available to soaring pilots and UAS a prescriptive threshold for maintaining percolation is identified that extends prior percolation analysis of soaring flight.

Analysis of contest flights suggests that pilot decisions are consistent with the objective of finding a thermal prior to falling below the percolation threshold, lending credence to the use of percolation value as a threshold for assessing and managing sporting risk.
The analysis also indicates that expert pilots typically exploit thermals prior to reaching this limit.
This is likely in an effort to obtain a more efficient path between points and may indicate a threshold at which pilots or autonomous systems should make greater course deviations or accept weaker thermals.

The larger dataset suggests that not only do pilots seek to stay ``percolating'' but that as they fall below the percolation threshold that they will accept slower climbs to achieve a percolating state once again.
As this behavior is observed in expert pilots it suggests that behaviors informed by explicitly tracking percolation are appropriate for pilot training and for use by UAS designed to similar objectives. \change{In practice, these insights could be implemented by computing and displaying the percolation value on the aircraft's navigation system. This ``percolometer'' would provide a risk measure to the pilot: if the percolation value falls below 1, the pilot should abandon a speed maximizing strategy and focus on finding a climb immediately\cite{Bird2019b}.}

When considering the weaker climbs at small percolation values it should be noted that in practice small percolation values are associated with low altitudes where thermals tend to be weaker in general\cite{Stull1988}.
In principle this could be a confounding effect on this analysis, however if the thermal intensity is defined for thermals of a particular strength then the percolation analysis still holds (and in fact would suggest very aggressively maintaining a value above 1 as ``strong thermals'' are less common at low altitudes).

This also suggests that percolation may offer a means to unify speed to fly theory and risk management. 
Often ``the'' MacCready number is used not only as a measure of the next thermal strength for speed optimization, but also as a proxy for risk management.
If the joint distribution of thermal spacing and strength can be determined, then the pilot can expect to be able to exploit a thermal whose spacing implies a percolation value greater than the threshold value.
This allows the MacCready value to be adjusted for constant risk with altitude, providing both speed to fly guidance as well as a threshold for the minimum thermal strength which should be accepted (the thermal strength whose spacing implies a percolation value at the threshold).
The result would be an adjustment of MacCready number with altitude, putting a mathematically rigorous underpinning to the approach proposed developed using heuristic value functions by Cochrane\cite{Cochrane1999}.

\section{Conclusions}

Graph percolation provides a useful measure of the risk of failing to find a thermal in cross-country soaring.
Analysis of flights conducted by human soaring pilots indicate that thermal exploitation typically begins while the altitude is great enough that the thermal graph should still percolate. 
Flight logs indicate that pilots accept slower climbs as the expected node degree decreases, an apparent response to risk of failing to find additional thermals.
These insights can be used to inform flight strategies for human pilots and autonomous systems seeking to extract energy from atmospheric motion.

\section*{Acknowledgments}
The author would like to acknowledge WeGlide for providing an anonymized set of flight data to support this research, as well as Daniel Sazhin who provided valuable insight into risk management and decision-making.

\bibliographystyle{ieeetr}
\bibliography{percolation.bib}

\appendix
\section{Percolation and Critical Node Degree}
Continuum percolation 

\section{Relating 1D and 2D Point Intensity}
\label{sec:intensity}
Continuum percolation typically considers spatially distributed random processes where the intensity is defined in events per unit area. The inter-thermal spacing is much easier to extract from flight data so a means to relate the spatial intensity and 1D intensity observed by the aircraft is needed.

For a 2D Poisson process, consider a disk of radius $r$ surrounding an arbitrarily chosen point. The probability mass function representing the number of thermals in the disk is given:
\begin{equation}
    P(n) = \frac{(\lambda_2 \pi r^2)^n\exp({-\lambda_2 \pi r^2})}{n!}
\end{equation}
evaluating this for $n=0$ gives the probability that no thermals lie within the disk:
\begin{equation}
    P(0) = \frac{(\lambda_2 \pi r^2)^0\exp({-\lambda_2 \pi r^2})}{0!}=\exp({-\lambda_2 \pi r^2})
\end{equation}
the complement of which gives the probability that at least one thermal is closer than $r$:
\begin{equation}
    P(n_{thermals} \ge 1) = 1 - P(0) = 1 - \exp({-\lambda_2 \pi r^2})
\end{equation}
which can be interpreted as the Cumulative probability that the distance to the nearest thermal is less than distance $r$:
\begin{equation}
    P(R \le r) = P(n_{thermals} \ge 1) = 1 - \exp({-\lambda_2 \pi r^2})
\end{equation}
The probability density for the distance to the nearest thermal is then the derivative with respect to distance of the previous equation:
\begin{equation}
    \begin{aligned}
        p(r) =& \frac{\partial P(R \le r)}{\partial r} \\
        =& \frac{\partial}{\partial r}(1 - \exp({-\lambda_2 \pi r^2})) \\
        =& 2\lambda_2 \pi r \exp({-\lambda_2 \pi r^2})
    \end{aligned}
\end{equation}
The expected value of this distribution can then be computed:
\begin{equation}
    \mathbb{E}(r) = \int_0^\infty r2\lambda_2 \pi r \exp({-\lambda_2 \pi r^2})dr=\frac{1}{2\sqrt{\lambda_2}}
\end{equation}

Because the Poisson process is memoryless, the expected distance distance from any point to a thermal is the same as the expected distance from one thermal to the next. 
If the distributions of thermals along a trajectory is also Poisson distributed, then the intensity of the process is one over the expected inter-thermal distance, so that:
\begin{equation}
    \frac{1}{\lambda_1} = \mathbb{E}(r) = \frac{1}{2\sqrt{\lambda_2}}
\end{equation}

Which completes the equivalence between one and two dimensional intensity, allowing the thermal density needed for percolation to be evaluated from the observed distribution of thermals along the aircraft's flight path.

\end{document}